\newcommand{\vy}[2]{#1_{\scriptscriptstyle #2}}
\def\hi{H\,{\sc i}} 
\def\hi{H\,{\sc i}} 
\def\ciii{C\,{\sc iii}} 
\def\civ{C\,{\sc iv}} 
\def\niv{N\,{\sc iv}} 
\def\nv{N\,{\sc v}} 
\def\oiv{O\,{\sc iv}} 
\def\ovi{O\,{\sc vi}} 
\def\ov{O\,{\sc v}} 
\def\neviii{Ne\,{\sc viii}} 
\def\siiv{Si\,{\sc iv}} 
\def\niii{N\,{\sc iii}} 
\def\pv{P\,{\sc v}} 
\def\svi{S\,{\sc vi}} 
\def\heii{He\,{\sc ii}} 
\def\siii{Si\,{\sc ii}} 
\def\cii{C\,{\sc ii}} 
\documentclass{emulateapj}
\shortauthors{Gabel et al.}
\shorttitle{VLT/UVES Observations of the AGN Outflow in HDFS Target QSO J2233-606}
\submitted{Accepted for publication in ApJ}

\begin{document}

\title{The AGN Outflow in the HDFS Target QSO J2233-606 \\
from a High-Resolution VLT/UVES Spectrum\altaffilmark{1}}
\author{Jack R. Gabel\altaffilmark{2}, Nahum Arav\altaffilmark{2},
Tae-Sun Kim\altaffilmark{3}}

\altaffiltext{1}{Based on public data released from the UVES commissioning
at the VLT/Kueyen telescope, ESO, Paranal, Chile.}
\altaffiltext{2}{Center for Astrophysics and Space Astronomy, University of Colorado, 389 UCB,
Boulder CO 80309-0389; jgabel@colorado.edu}
\altaffiltext{3}{Institute of Astronomy, Madingley Road, Cambridge CB3 0HA, UK}

\begin{abstract}
  We present a detailed analysis of the intrinsic
UV absorption in the central HDFS target QSO J2233-606, based on a high-resolution,
high S/N ($\sim$25 -- 50) spectrum obtained with VLT/UVES.
This spectrum samples the cluster of intrinsic absorption 
systems outflowing from the AGN at radial 
velocities $v \approx -$5000 -- 3800 km~s$^{-1}$ in
the key far-UV diagnostic lines - the lithium-like CNO doublets and 
\hi\ Lyman series.
We fit the absorption troughs using a global model of all detected lines
to solve for the independent velocity-dependent covering factors of the continuum
and emission-line sources and ionic column densities.
This reveals increasing covering factors 
in components with greater outflow velocity.
Narrow substructure is revealed in the optical depth profiles, 
suggesting the relatively broad absorption is comprised of
a series of multiple components.
We perform velocity-dependent photoionization modeling, which
allows a full solution to the C, N, and O abundances, as well
as the velocity resolved ionization parameter and total column density.
The absorbers are found to have supersolar abundances,
with [C/H] and [O/H] $\approx$0.5 -- 0.9, and [N/H] $\approx$ 1.1 -- 1.3, 
consistent with enhanced nitrogen production expected from secondary 
nucleosynthesis processes.
Independent fits to each kinematic component give 
consistent results for the abundances.
The lowest-ionization material in each of the strong absorbers
is modeled with similar ionization parameters.  
Components of higher-ionization (indicated by stronger \ovi\ relative to
\civ\ and \nv) are present at velocities 
just redward of each low-ionization absorber.
We explore the implications of these results for the 
kinematic-geometric-ionization structure of the outflow.
\end{abstract}


\section{Introduction}

   Active galactic nuclei (AGN) often show mass outflow as blueshifted 
absorption in their restframe UV and X-ray spectra.
Many recent studies have explored the potential global effects of AGN outflows
on all scales of their environment: from influencing the growth of the
central supermassive black hole \citep{blan99,blan04}, the
evolution of the host galaxy \citep{silk98,scan04},
the luminosity function of quasars \citep{wyit03}, and the magnetization
and chemical enrichment of galactic gas \citep{furl01} and
the intergalactic medium \citep{cava02}.
It is unclear if the outflows seen as UV/X-ray absorption are related
to these global phenomena, or if they are connected to other AGN components 
showing evidence for bulk outflow motion such as 
the relativistic jets seen in radio observations and the 
emission-line gas.

   Measured ionic column densities provide the basis for physical 
interpretation of AGN outflows.
Detailed UV spectral studies have shown measurements 
of these crucial parameters are often not straightforward.
Analyses of absorption doublets and multiplets show the absorbers
typically only partially occult the background emission sources 
\citep[e.g.,][]{hama97,barl97},
with potential complex scenarios involving different covering factors
for different background emission sources \citep{gang99,gabe03}, 
velocity-dependent covering factors \citep[e.g.,][]{arav99},
and possibly inhomogeneous distributions of absorbing material \citep{deko02a,arav05}.
Detailed spectral studies of rich absorption line spectra at high
spectral resolution and high signal-to-noise (S/N) are needed to deconvolve
these effects from the observed absorption lines, and thereby determine
the total column density, ionization, and chemical abundances of the outflow 
via modeling.

   The quasar absorption-line spectrum of QSO J2233-606 (hereafter J2233) has been 
intensively observed and studied due to its location in the center of the 
Hubble Deep Field South (HDFS; Ferguson 1998), its high redshift 
($z_{em}$=2.238), and relative brightness ($B=$17.5) 
\citep{seal98,sava98,outr99,proc99,cris00,kim01}.  
In addition to the rich intervening absorption spectrum, these observations have 
revealed several ``associated" absorption systems 
\citep{seal98,sava98,outr99,peti99}.
These have been interpreted as outflowing material associated with the 
QSO in J2233 due to their partial line-of-sight coverage of the compact
AGN emission, relatively broad widths, 
and proximity in redshift to the host galaxy ($z_{abs} \approx$2.198 -- 2.21).
\citet{peti99} presented an analysis of the (rest-frame) UV
associated absorption in J2233 using high-resolution optical spectra
and supplementary long observations with the {\it Hubble Space
Telescope} ({\it HST}) obtained as part of
the HDFS program, which sampled lines spanning a large range of ionization.
Their photoionization modeling of one kinematic component 
($z_{abs}=$2.198) showed that in addition to a relatively 
low-ionization region giving 
the \hi, \civ, and \nv\  absorption, a high-ionization 
zone is needed to reproduce the strong \neviii\  absorption.
They found this component was consistent with having approximately 
solar metallicity, with qualitative evidence for the N/C abundance 
ratio enhanced relative to the solar value.
They also found ion-dependent covering factors and suggested this may be due
to a smaller covering factor of the emission-line region than the 
continuum source \citep[see also][]{outr99}.

     J2233 was observed for a total of 14 hours with VLT/UVES 
as part of a large program to study intervening systems in quasar
absorption spectra \citep{kim01}.
We present here a detailed analysis of the absorption-line properties and physical
conditions in the associated absorption systems based on this spectrum.
Our study is outlined as follows: in \S 2, we review the observational
details and present the intrinsic absorption spectrum from these observations.
In \S 3, we make use of the broad spectral coverage which includes the CNO lithium-like
doublets and \hi\  Lyman series lines at high resolution and high S/N to
solve for the individual covering factors of the emission-line
and continuum sources and the velocity-dependent ionic column densities.
In \S 4, we incorporate velocity-dependent photoionization modeling of
each kinematic component to highly constrain the models and derive the
full solution to the physical conditions (total column density and ionization)
and chemical abundances of the outflow.
In \S 5, we explore the implications of this analysis for the outflow's 
geometric, kinematic, and ionization structure.

\section{VLT/UVES Observations and the Intrinsic Absorption in J2233}

   J2233 was observed with VLT/UVES for a total of 50.4 ks 
between 1999 October 8 -- 16 during the Commissioning 
of UVES (we refer the reader to Cristiani \& D'Odorico 2000 and
Kim et al. 2001 for full details of the observations).
These observations used the dic1 346$\times$580 (28.8 ks) and 
dic2 437$\times$860 (21.6 ks) settings of UVES, giving full 
spectral coverage between 3060 -- 10000 \AA.
The data were processed with the ESO-maintained MIDAS
ECHELLE/UVES package, as described in \citet{kim01}.
The reduced spectrum was resampled to 0.05 \AA\  bins and
normalized locally using a 5th order polynomial fit.
The resulting final spectrum used in the analysis had
resolving power ranging from $R \approx$ 30,000 -- 50,000 and S/N between 
$\sim$ 25 -- 50 for the spectral regions of interest, with
increasing values towards longer wavelength.

   The UVES spectrum covers the rest wavelengths of a host of
key diagnostic far-UV lines for intrinsic absorbers,
including the CNO lithium-like doublets 
(\ovi\ $\lambda\lambda$1032,1038, \nv\ $\lambda\lambda$1238,1242, 
\civ\ $\lambda\lambda$1548,1551) and Lyman series lines which are ubiquitous
in intrinsic UV absorption spectra, as well as
a number of less common lines that probe lower ionization
(e.g., \ciii\ $\lambda$977, \siiv\ $\lambda\lambda$1393,1403, 
\niii\ $\lambda$991) and lower abundances (e.g., \svi\ $\lambda\lambda$933,944, 
\pv\ $\lambda\lambda$1118,1128).
This is the first high-resolution spectrum of J2233
that spans all these lines with the same instrument 
and in the same epoch.
The UVES spectrum is also of significantly 
higher quality than the previous observations:
it has about twice the resolution and
4 -- 5 times higher S/N at the \civ\ lines,
twice the S/N at \nv\ and Ly$\alpha$,
and six times better resolution and 2 -- 5 times
higher S/N than the HST/STIS G430M grating 
observation of the \ovi/Ly$\beta$ spectral
region (see Table 1 in Petitjean \& Srianand 1999).

\begin{deluxetable}{llll}
\tablewidth{0pt}
\tablecaption{Intrinsic Absorption Components in QSO J2233-606 \label{tbl-1}}
\tablehead{
\colhead{Component} & \colhead{Velocity\tablenotemark{a}} & \colhead{FWHM\tablenotemark{a}}  &
\colhead{Lines\tablenotemark{b}} \\ 
\colhead{} & \colhead{km s$^{-1}$} & \colhead{km s$^{-1}$} & \colhead{} }
\startdata
1 & -4805 & 56  & \ovi, \nv, \civ, Ly$\alpha$, Ly$\beta$ \\
2 & -4137 & 171 &  \ovi, \nv, \civ, Ly$\alpha$, Ly$\beta$   \\
3 & -3971 & 131 &  \ovi, \nv, \civ, Ly$\alpha$\tablenotemark{c}  \\
4 & -3830 & 55  & \ovi  \\
5 & -3705 & 50  & \ovi  \\
6 & -3634 & 32  & \ovi  \\
\enddata
\tablenotetext{a}{~Centroid velocity and FWHM are for \civ\  $\lambda$1548 
for components 1 -- 3 and \ovi\ $\lambda$1032 for components 4 --6.}
\tablenotetext{b}{~Detected lines in each kinematic component in VLT/UVES spectrum.}
\tablenotetext{c}{~Ly$\beta$ contaminated with Ly$\alpha$ forest.}
\end{deluxetable}

    Figure 1 shows the associated absorption spectra for several
key diagnostic lines in the UVES spectrum, plotted on velocity scale with 
respect to the restframe of the AGN. 
We also include the \neviii\ $\lambda\lambda$770,780 doublet observed 
with the Space Telescope Imaging Spectrograph (STIS) using the E230M 
grating as part of the HDFS observations \citep{ferg98}.
Figure 1 shows three strong absorption components centered 
at $v_r \approx -$4800, $-$4150, and $-$3950 km~s$^{-1}$ (hereafter 
components 1, 2 and 3, respectively), identified as associated 
systems in earlier observations based on the partial line-of-sight covering
factors inferred from the doublet lines \citep{outr99,sava98,peti99}.
They are also distinguished from intervening absorbers in the 
high-resolution of the UVES spectrum by their relatively large widths, 
as seen for example by comparison with the Ly$\alpha$ forest lines in Figure 1.
These absorbers exhibit absorption, in increasing strength,  
in the \civ, \nv, and \ovi\  doublets; they are 
not detectable in the lower ionization lines \ciii\  or the 
\siiv\ doublet. 
Though the STIS spectrum has substantially lower S/N and there is
potential contamination from intervening absorption, \neviii\  
also appears strong in components 1 and 3, and possibly component 2.
Ly$\alpha$ is present in all three components, and
weaker than in the CNO doublets; Ly$\beta$ is detected
in component 1, with a potential weak notch
present in component 2, while component 3 is contaminated
with intervening Ly$\alpha$ forest absorption. 
The higher order Lyman lines are in a low S/N region of the UVES spectrum,
but there is no detectable absorption in these lines 
within the limits of the noise.
There are three additional relatively weak systems seen in the 
\ovi\  doublet (and  possibly \neviii) just redward of component~3
\citep[components 4 -- 6 at $v_r \approx -$3850, $-$3700, and 
$-$3625 km~s$^{-1}$;][]{peti99}.
These systems are not detectable in absorption in any other lines. 
Table 1 summarizes the kinematic properties and lines present in each 
kinematic component.

\begin{figure}
\includegraphics[width=8.5cm]{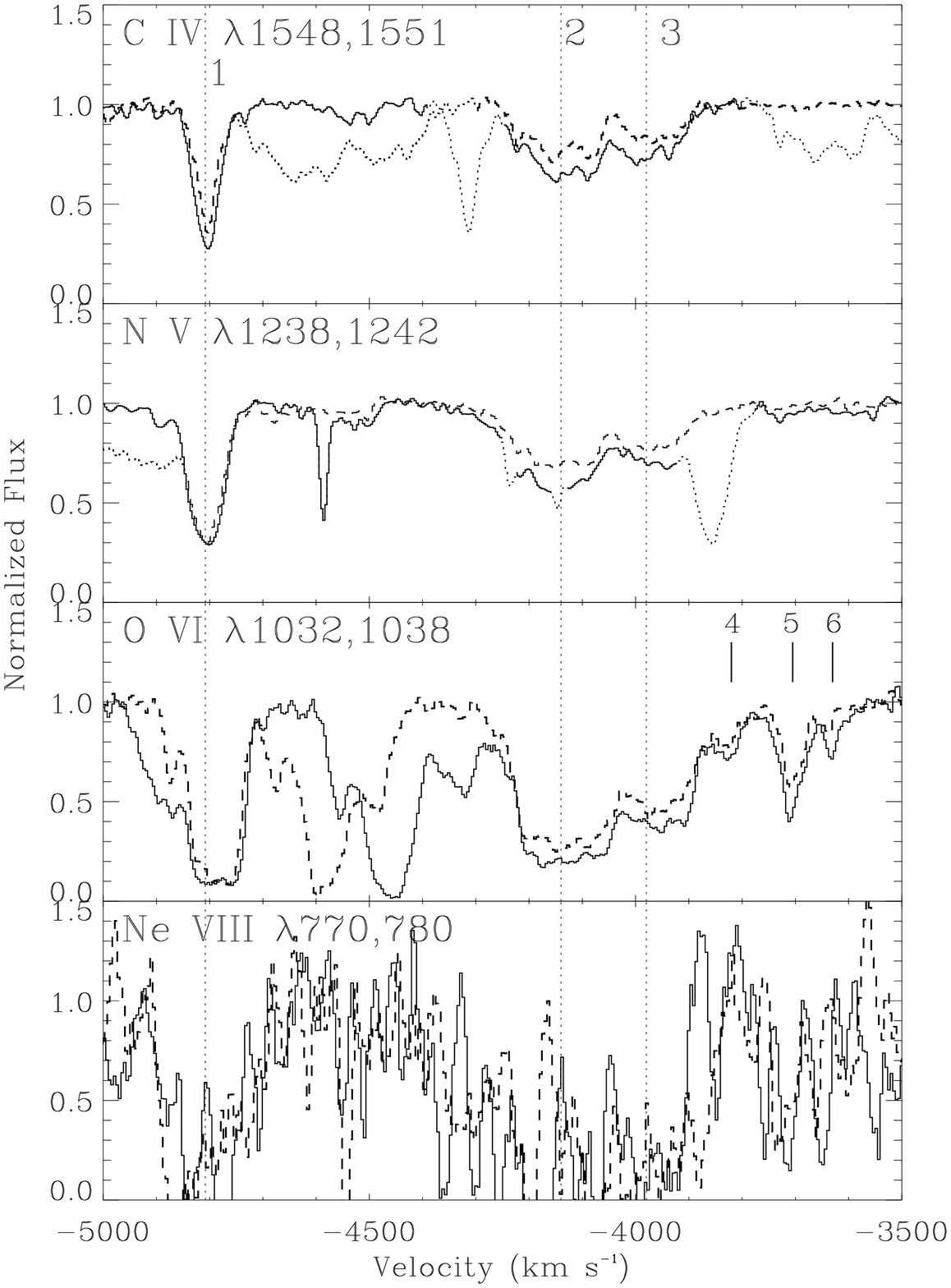}
\vspace*{0.35 in}
\caption{Normalized absorption profiles from VLT/UVES spectrum of QSO 
J2233-606.  Each line is plotted as a function of radial velocity
with respect to the emission-line redshift of the quasar ($z=$2.238).
The \neviii\ doublet from the HDFS E230M observations is also included.
Long wavelength doublet members and Ly$\beta$ are plotted with dashed
lines, short wavelength doublet lines\label{fig1a}}
\end{figure}

\addtocounter{figure}{-1}

\begin{figure}
\includegraphics[width=8.5cm]{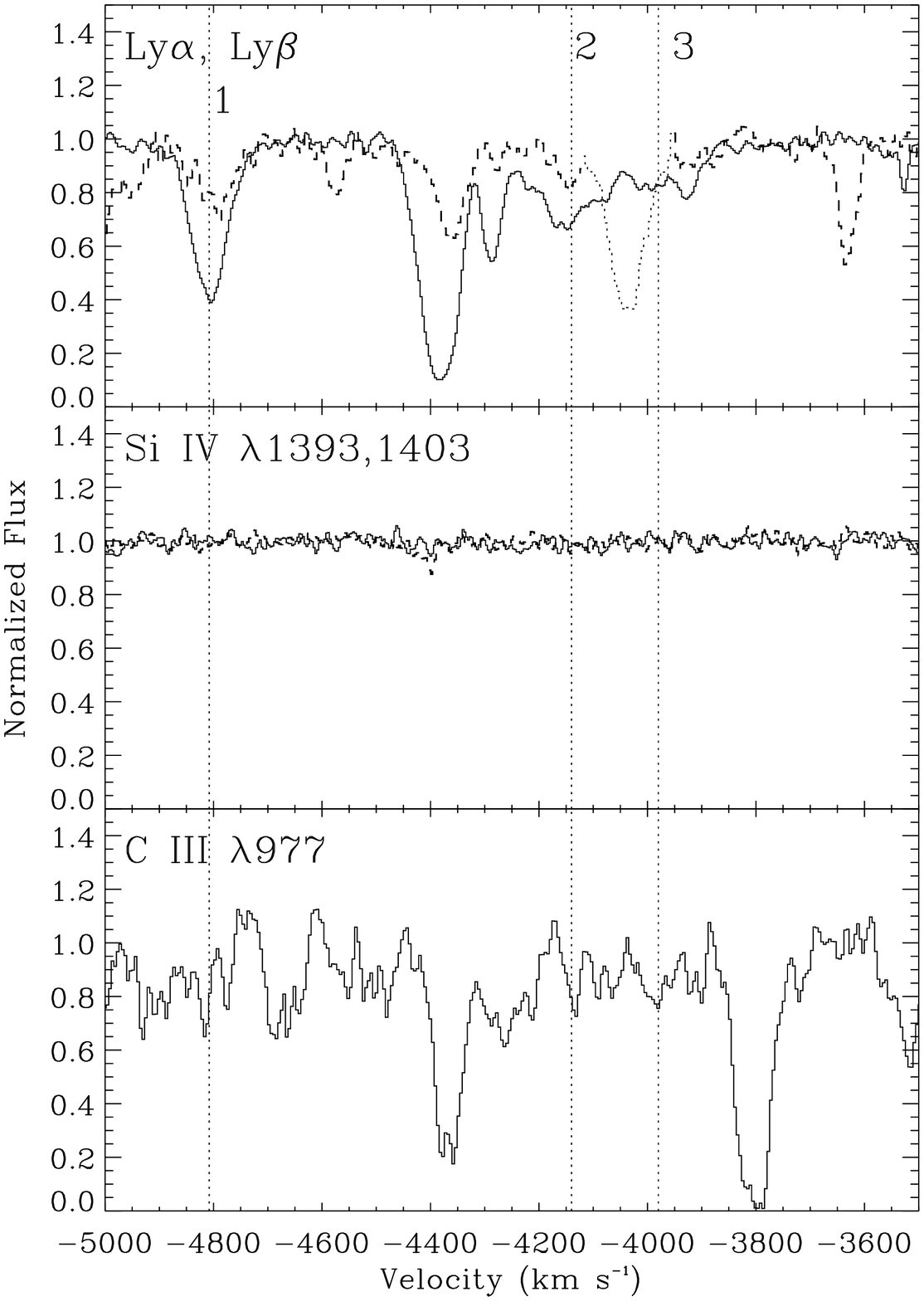}
\vspace*{0.35 in}
\caption{ and Ly$\alpha$ with solid lines. 
Regions plotted with dotted lines indicate contamination 
by intervening
absorption or the other doublet member.  
Dotted vertical
lines denote the three strong kinematic components analyzed in this
study; tick marks show the three additional components seen only
in \ovi\  and \neviii.  \label{fig1b}}
\end{figure}  

\newpage

\section{Velocity-Dependent Ionic Column Densities
and Covering Factor Solutions}

\subsection{Fitting Formalism}

  The observed intrinsic absorption troughs 
are a convolution of line-of-sight covering factor ($C$) and optical
depth ($\tau$).
These factors must be deconvolved in order to obtain ionic column densities
for modeling the ionization and abundances of the outflows; 
additionally, the covering factors derived from this separation
give important constraints on the absorption-emission geometry.
With the two lines of an absorption doublet, single values
for $C$ and $\tau$ can be derived at each velocity 
\citep[e.g.,][]{hama97,barl97}; 
however, complex structure in the absorber
\citep[e.g.,][]{deko02a,arav05} or the 
background emission sources \citep[e.g.,][]{gang99,gabe03} 
requires more than two lines to obtain the full solution.
In particular, if continuum and line emission both underly the absorption
features, the individual emission sources will in general have different 
covering factors, with implications for the doublet solution and relative values
between different lines \citep[see \S 3 in][]{gabe05a}.

   We applied a global fitting approach to determine the velocity-dependent covering
factors and column densities of the outflow in J2233, following the method 
outlined in \citet{gabe05a}.  This approach allows for 
treatment of different covering factors for the continuum and emission-line 
sources by simultaneously fitting the observed 
absorption lines from multiple ions. 
The general absorption equation to solve for the normalized intensity of the $j^{th}$ line is:
\begin{equation}
I^j = \Sigma_i  [ R_i^j  (C_i^j  e^{-\tau^j} + 1 - C_i^j) ], 
\end{equation}
where the $i^{th}$ individual emission source contributes
a fraction $R_i^j = I_i^j / \Sigma_i [I_i^j]$ to the total
intrinsic intensity and has covering factor $C_i^j$ \citep{gang99,gabe05a}.
Here we have assumed that a single, uniform column density for each ion 
occults all emission sources.  The effective covering factor for each 
line is the intensity-weighted combination of the individual covering factors,
\begin{equation}
C^{j} = \Sigma_i C_i^j R_i^j.
\end{equation}

    For components 1 -- 3 in the J2233 outflow, there are up to 8 detected 
lines available to constrain the absorption parameters (fewer where lines 
are contaminated with other absorption): the C, N, O doublets, Ly$\alpha$, 
and Ly$\beta$.  We assumed all ions share the same covering factors of the 
continuum and emission line sources ($C_c$ and $C_l$), and simultaneously
fit the individual covering factors and \civ, \nv, \ovi, and \hi\  column densities.  
These parameters were solved in each velocity bin of the absorption profiles
using equation 1 via the Levenberg-Marquardt non-linear least-squares minimization 
technique \citep[additional details are given in][]{gabe05a}.  
We limited our search of the covering factor parameter space to
the physically meaningful range,  0$\leq C_i \leq$1.  
Regions of contamination with other absorption were omitted from
the fitting. This included the far blue wing of component 1 \nv\ $\lambda$1242 
and red wing of component 3 \nv\ $\lambda$1238 due to blending of these lines, and
obvious intervening Ly$\alpha$ forest lines: at $v \approx-$4240 and 
$-$4140~km~s$^{-1}$ in \nv\ $\lambda$1238 and $v \approx-$4030~km~s$^{-1}$ 
in Ly$\beta$ (see Figure 1).

   A source of modeling uncertainty is the intrinsic flux levels of the continuum
and BLR emission; the VLT/UVES spectrum is not flux-calibrated and
thus the intrinsic shape of the AGN emission cannot be determined from these data.
Therefore, to determine the relative contributions, $R_c$ and $R_l$, we adopted 
results from previous observations of J2233.
For \ovi\  and Ly$\beta$, we used the HDFS STIS G430M spectrum obtained in 1998 October.  
For the longer wavelength lines, we used published results from earlier
ground-based observations:  Ly$\alpha$ and \nv\  were 
from data in \citet{outr99} obtained between 1997 August -- 1998 August 
and \civ\  from a spectrum in \citet{sava98} obtained in 1997 October.
Based on these published results, our adopted contributions from the BLR 
for fitting were: $R_l \approx$ 0.15 for \ovi\ ; 0.3 for \civ\ ; 0.4 for \nv\ and
Ly$\alpha$; and 0.1 for Ly$\beta$.
A caveat is that since these observations were obtained at previous epochs (the
UVES spectrum was obtained in 1999 October),
there may have been variability in the intrinsic emission that could affect our
results.
However, we believe this effect will be minimal since 
high-luminosity QSOs like J2233 do not typically exhibit strong 
variability over timescales of a few years.
For example, in a recent analysis of the $\approx$2500 QSOs in the SDSS database having
multiple observations separated by more than 50 days,
\citet{whil05} found only 12\% exhibited significant 
variability.
Additionally, {\it HST}/STIS observations of J2233 obtained a year apart 
had identical fluxes \citep[see Figure 2 in][]{peti99}.

\subsection{Results}

    The results of the fits for the covering factors and optical depths 
are shown in Figures 2 and 3, respectively.  
The plotted error bars represent the formal 1$\sigma$ statistical errors in 
the best-fit parameters, corresponding to values giving $\Delta \chi^2 =$1, 
and were computed from the diagonal elements of the 
covariance matrix for the optimal fit (Bevington 1969); they 
do not include any uncertainties associated with the intrinsic flux levels.
For cases where the covering factor solution is a boundary 
value (0,1), no covariance matrix elements are computed since it is 
not a minimum in the solution; for these cases, we estimated uncertainties 
by deriving solutions for models keeping the parameter fixed, and
finding the value giving $\chi^2=$1 from the best-fit solution at the
boundary value.
For $\tau \gtrsim$3, an absorption line can be considered to be
saturated -- for these large optical depths, changes in $\tau$ affect 
the absorption depths less than the uncertainties in the flux levels. 
Thus, we adopt solutions giving $\tau \geq$3 as lower limits of 3. 
This occurs over much of the core of the component 1 \ovi\  profile,
and over a narrower region of the core of \nv\  in this component.
The fits in several bins in \ovi\  in components 2 and 3 also reach
these large values of $\tau$, mostly corresponding to the peaks in the
$\tau$ profiles in the other ions.  
In Figure 4, the absorption profile 
fits (red), derived by inserting the best-fit solutions into equation 1, 
are compared with the observed profiles. 
Most of the absorption in all lines is fit well by this model.    
In \S 4.5, we explore some of the minor discrepancies in the fit and their 
implications for our fitting assumptions and subsequent analysis.

\begin{figure}[location=t]
\includegraphics[angle=90,width=9cm]{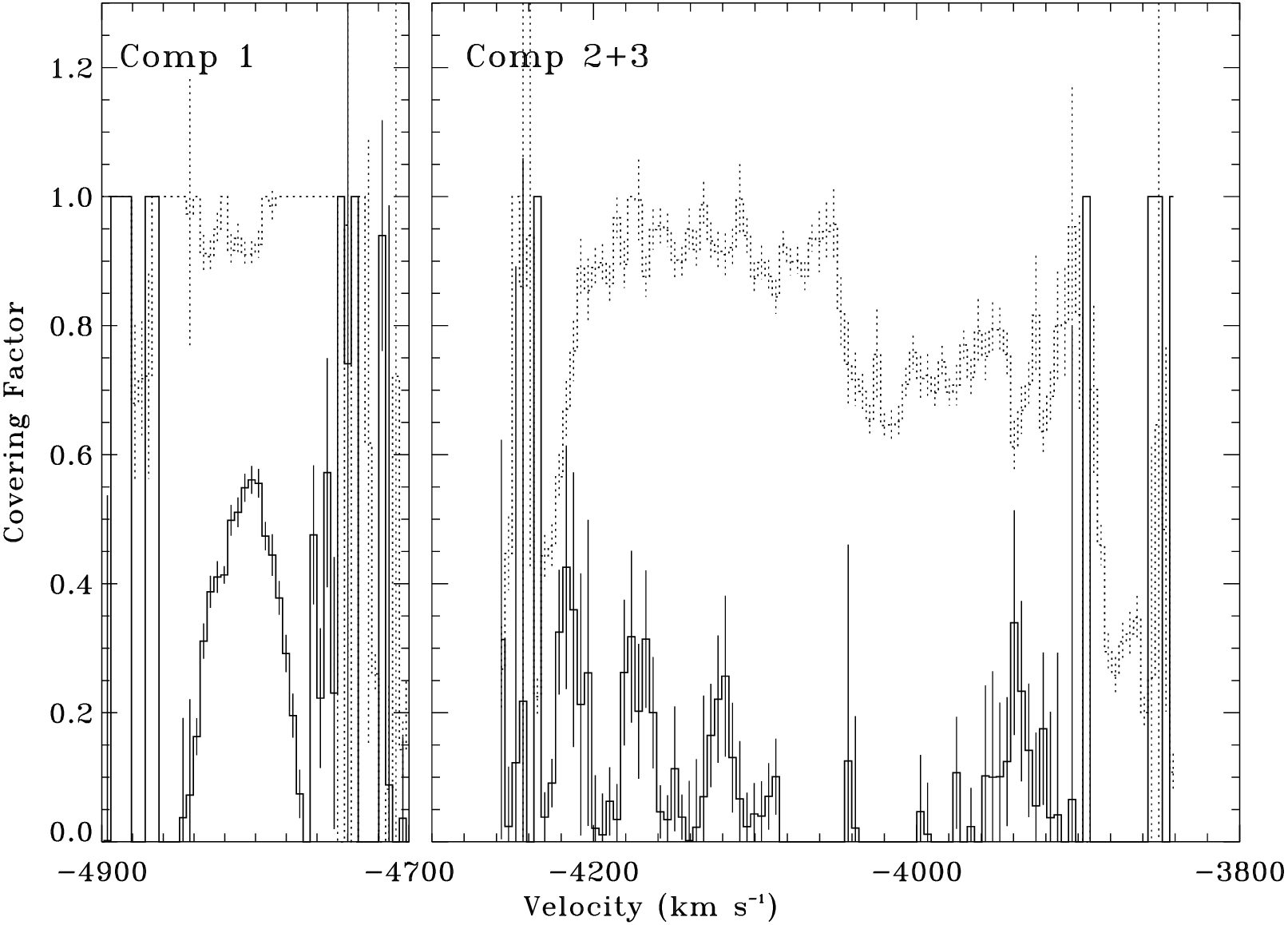}
\vspace*{0.1 in}
\caption{Global-fit solution to the line-of-sight absorption covering factors.
 Broad emission-line region covering factors are plotted with solid histograms,
continuum covering factors with dotted histograms. Solutions were obtained  
by simultaneously fitting all detected lines in the UVES spectrum, assuming all 
ions have the same covering factors (see text). \label{fig2}}
\end{figure}

   Figure 2 shows the continuum source covering factor is near unity for all 
velocity bins in components 1 and 2.
Component 3 only partially occults the continuum source, with 
$C_c \approx$0.75. 
The BLR emission is partially occulted in component 1, peaking at 
$C_l \approx$0.55 in the core and decreasing sharply in the wings of 
the profile.
In components 2 and 3, the BLR coverage is low, with $C_l=$0 at better
than the 2$\sigma$ level at nearly all velocities.
Results are summarized in Table 2.
The overall covering factor solutions 
give compelling support for the assumptions made in the calculations.  
Since each velocity bin provides an independent solution, 
the similarity in the  
individual $C_c$ solutions across each kinematic 
component supports the distinction of the continuum and BLR 
coverage as a major factor in the formation of the observed troughs.  
If this were instead an artificial distinction, it seems unlikely 
that the same result for $C_c$ would be obtained over 
the numerous bins associated with each component, 
particularly since they exhibit a range of optical depths 
for each ion and optical depth ratios between ions.
These results are also consistent with the general picture of the AGN emission
geometry, in which the BLR is much more extended than the compact UV continuum
source (further implications of these results are treated in \S 5).

\begin{figure}
\includegraphics[width=8.5cm]{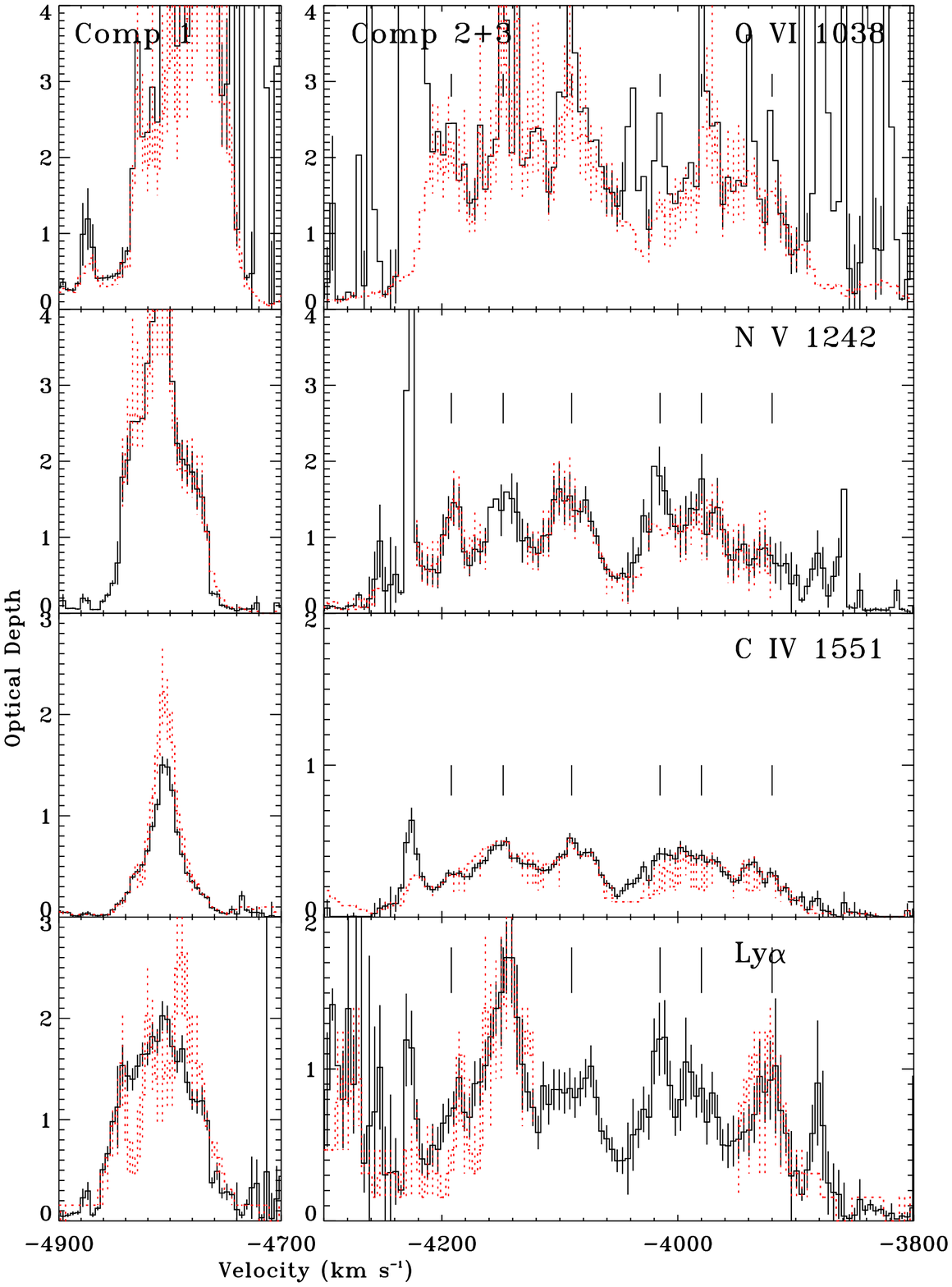}
\vspace*{0.35 in}
\caption{Solutions to the intrinsic absorption optical
depth profiles. Global-fit solutions are shown with black histograms. 
Solutions to line pairs for each ion individually, assuming $C_c$ from the 
global-fit solution, are shown as red histograms (see text).  
For the doublets, $\tau$ for the red lines are shown. Tickmarks
denote peaks in the narrow substructure seen in the solution to 
components 2 \& 3.\label{fig3}}
\end{figure}

    In Figure 3, interesting substructure in the optical depth profiles
is revealed in components 2 and 3.  The relatively broad features break 
down into a series of narrow subpeaks in $\tau$, with coincident
kinematic structure seen in the different ions.  We note some of the 
substructure in the optical depth profiles coincides with the fluctuations 
in the $C_l$ solution in Figure 2.
Thus, to test if the structure in the $\tau$ solutions is an
artifact of the minimization fitting, 
we also fit the absorption constraining $C_l=$0 at all velocities.  
Resulting fits to the profiles are plotted in Figure 4 (dotted red 
histograms), showing similarly good fits over most of the absorption.  
The resulting $\tau$ profiles of this fit show similar overall structure 
as the solutions shown in Figure 4 with $C_l$ a free parameter,
indicating this structure is likely real. 
Component 1 is seen to have velocity-dependent
ionization structure; \civ\  and \nv\  exhibit a narrow peak at
$v \approx-$4810~km~s$^{-1}$, in contrast with \ovi, which
is strong (saturated) over a broader range of velocities
extending redward to $-$4740~km~s$^{-1}$. 
\hi\  lacks the sharp narrow peak in the core, exhibiting
a relatively broad base of absorption over the profile.
We integrated the optical depth profiles over velocity, and computed 
total ionic column densities associated with each kinematic component.
Results are summarized in Table 2.

\begin{deluxetable}{lrrrrrrr}
\tablewidth{0pt}
\tablecaption{Integrated Absorption Parameters \label{tbl-2}}
\tablehead{
\colhead{A\tablenotemark{a}} & \colhead{$N_{{\rm O}\,{\rm\sc VI}}$\tablenotemark{b}} & \colhead{$N_{{\rm N}\,{\rm\sc V}}$}  & \colhead{$N_{{\rm C}\,{\rm\sc IV}}$}  &
\colhead{$N_{{\rm H}\,{\rm\sc I}}$} & \colhead{$N_{{\rm C}\,{\rm\sc III}}$} & \colhead{$C_c$\tablenotemark{c}} &
 \colhead{$C_l$\tablenotemark{c}}}
\startdata
1\tablenotemark{d} & $\geq$13 & $\geq$7  & 1.4  & 1.1  & $\leq$0.1 & 0.95 -- 1 & 0.55   \\
2 & 19 & 6.5  & 1.4  & 1.0  & $\leq$0.1  & 0.9 -- 1 & 0 -- 0.2  \\
3 & 15 & 5.7  & 1.2  & 0.8  & $\leq$0.1 & 0.65 -- 0.75 & 0 -- 0.2  \\
\enddata
\tablenotetext{a}{~Intrinsic absorption kinematic component (see Table 1).}
\tablenotetext{b}{~All column densities, $N_{ion}$, are in units 10$^{14}$ cm$^{-2}$; 
\ion{O}{6}, \ion{N}{5}, \ion{C}{4}, and \ion{H}{1} were measured from the global-fit 
solutions (\S 3).}
\tablenotetext{c}{~Derived continuum ($C_c$) and emission-line ($C_l$) covering factors at the core
velocities of each component.}
\tablenotetext{d}{~Heavy saturation in \ovi\  at $v \approx -$4800 -- $-$4750;
\nv\  saturated at $v \approx -$4820 -- $-$4790.}
\end{deluxetable}

  Our measurements can be compared with those of \citet{peti99}, 
who present results only for component 1.
The \hi\  measurements are similar (within $\approx$20\%), however, our
\civ\  column density is only about half their measured value, while
\nv\  is at least two times greater, with saturation at some velocities. 
Additionally, we find \ovi\  is heavily saturated over much
of the profile (see Figures 1 and 3) giving only a lower limit on its
column density; this was not discernible in the {\it HST}/STIS G430M spectrum 
used in the \citet{peti99} study due to the limited S/N.

\section{Derivation of the Abundances, Ionization, and Total Column
Density in the Outflow}

\subsection{Modeling Methodology and Input Parameters}

  To determine the physical conditions and abundances in the outflow in J2233,
we compare the measured ionic column densities derived in \S 3 with predictions from 
models computed with the photoionization code Cloudy \citep{ferl98}.
A model is specified by the spectral energy distribution (SED) of the
ionizing continuum, the total hydrogen column density ($N_H$) and
elemental abundances in the absorber, and the ionization parameter 
($U = Q / 4 \pi R^2 n_H c$), which gives the ratio of the density
of H-ionizing photons at the face of the absorber to the gas number density, $n_H$.

     In components 1 -- 3 in the J2233 outflow, model constraints are 
available from measurements of \civ, \nv, \ovi, 
and \hi.  To fully model these four ions, five parameters
must be specified for the adopted SED: $U$, $N_H$, and the abundances 
of C, N, and O relative to H, thus the solution for a single measurement 
of these ions is underconstrained.  However, the high-resolution of the 
column density solutions derived in \S 3 can be utilized to give 
additional model constraints thereby allowing the full solution 
of the absorber properties.
In the most straightforward model, the elemental abundances will be the
same throughout the absorber; thus we model the column densities measured in 
{\it each velocity bin} of an absorption component and link the solutions
for all velocities by requiring the C, N, and O abundances to be the same at
all velocities. 
Given the highly resolved absorption profiles, 
this greatly increases the number of constraints for the models, 
which allows the fitting of the abundances and 
gives the {\it velocity-dependent} solution of the physical
conditions ($N_H$ and $U$) in individual absorption
components.  Another advantage of this method compared with modeling
integrated column densities over a component is that regions of line
saturation can be isolated.

\begin{figure}
\includegraphics[width=8cm]{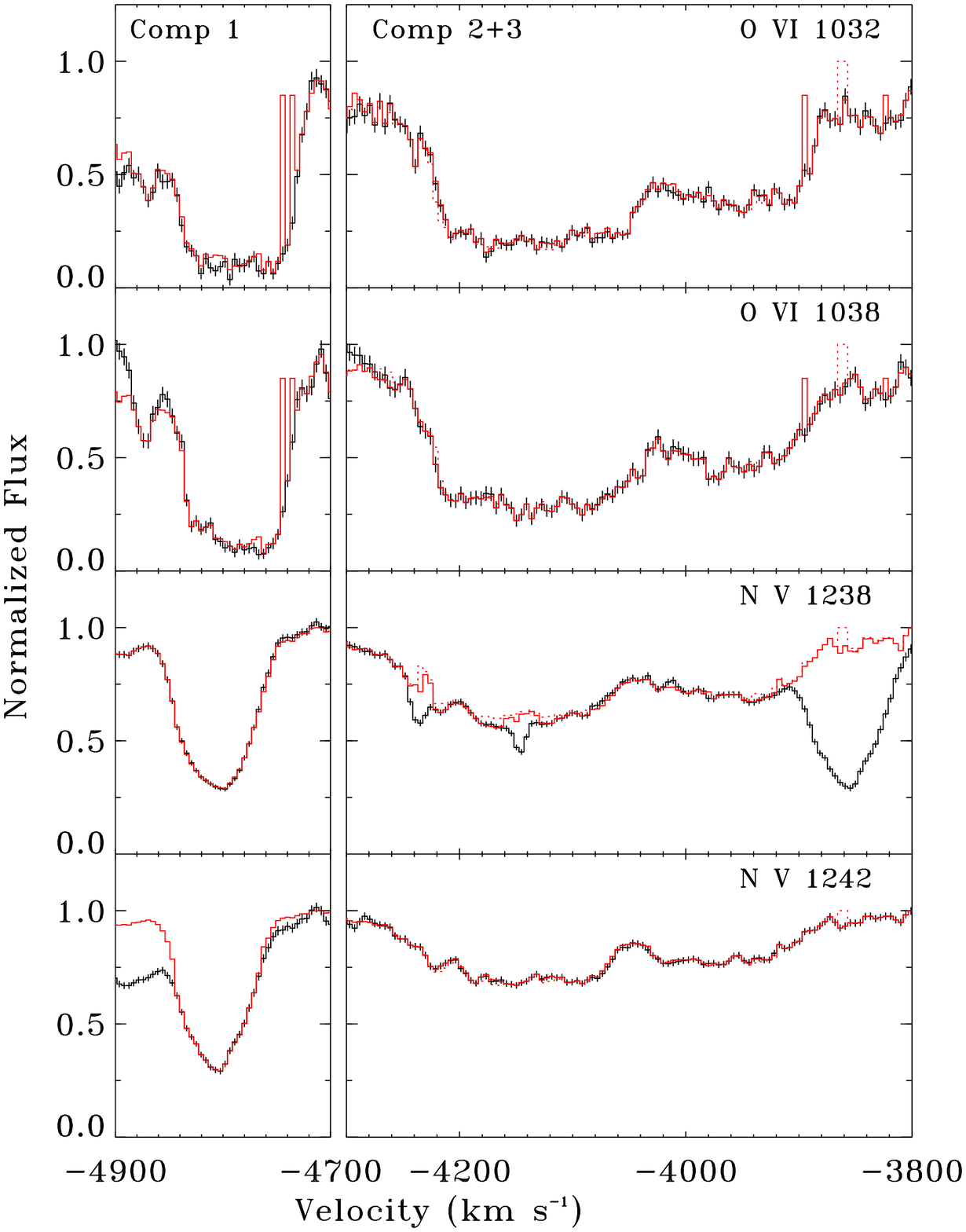}
\vspace*{0.35 in}
\caption{Best-fit intrinsic absorption profiles for the eight lines 
solved with the global-fitting to $\tau$ -- $C$.  Model profiles (solid 
red histograms) were derived from the best-fit covering factors and column densities
show in figures 2 and 3, and are plotted with the observed normalized
profiles (black).\label{fig4a}}
\end{figure}

\addtocounter{figure}{-1}

\begin{figure}
\includegraphics[width=8cm]{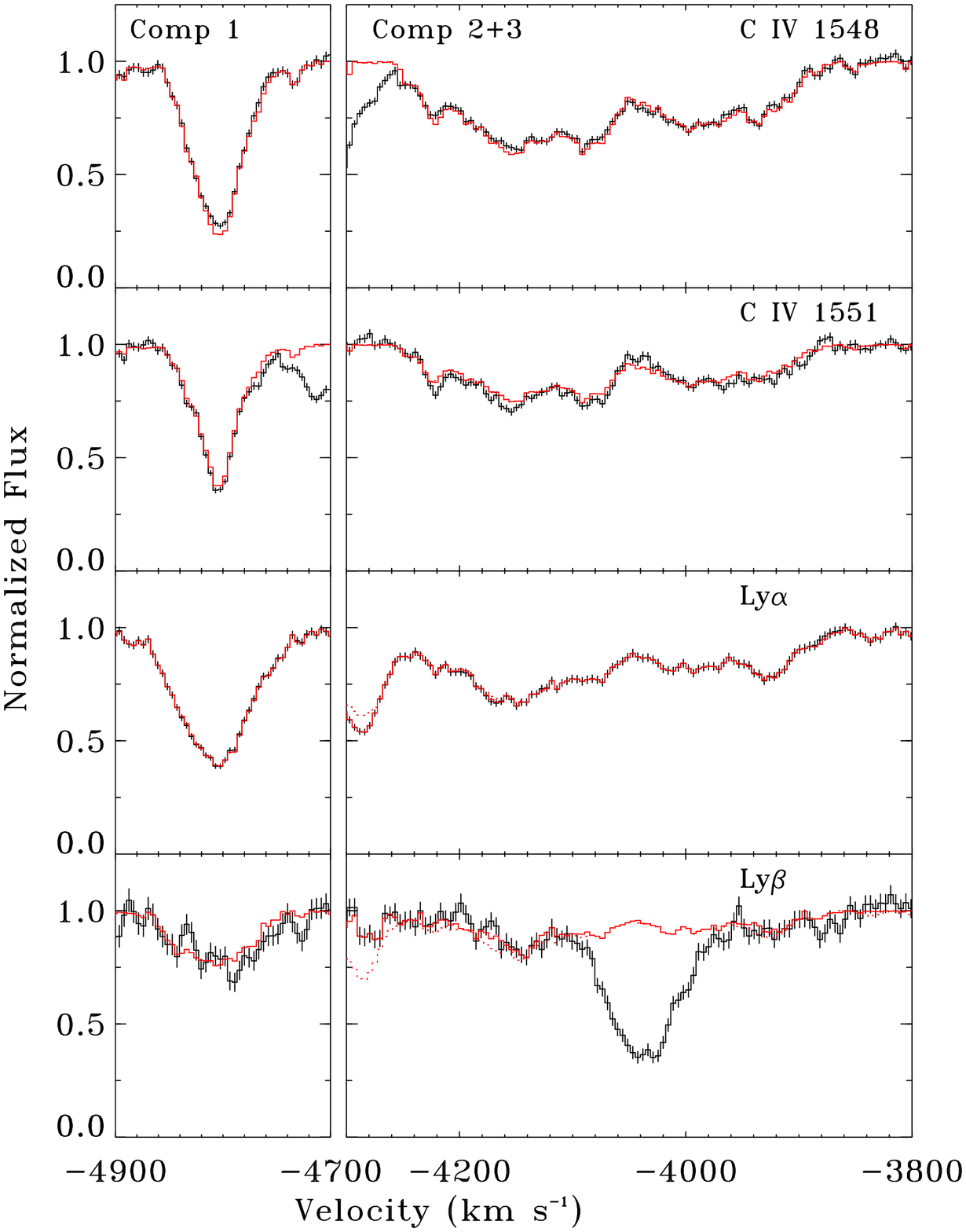}
\vspace*{0.35 in}
\caption{Model profiles for $\tau$ -- $C$ solutions in which the BLR
covering factor was fixed at $C_l =$0 are shown for components 2 and 3 (dotted 
red), and are seen to be nearly identical as the solutions with $C_l$ a free 
parameter.  \label{fig4b}}
\end{figure}

  We adopted the SED used in \citet{gabe05b} for our modeling, which
consists of a broken power law ($F_{\nu} \propto \nu^{-\alpha}$) in the
ionizing continuum, with $\alpha =$ 1.4 and 0.7 over energy intervals
0.0136 -- 0.6 and 0.6 -- 50 keV, respectively.
The ionizing continuum used in the calculations is a source of uncertainty;
a study by \citet{laor99} shows that the SED for quasars may be somewhat steeper than  
the one adopted for our modeling, which is typical for lower luminosity AGN.
We have done extensive testing with different SEDs to explore its effect 
and find it does not significantly change our main results on the ionization and
abundances in the outflow.

     We computed a grid of photoionization models in $U$, $N_H$ for 
solar elemental abundances \citep{grev89}.
We then fit the absorbers 
by minimizing the $\chi^2$ value in comparing the measured ionic column densities
with the grid of model predictions.
The $U$, $N_H$ associated with each of the $n_v$ velocity bins 
are free parameters, with all velocity bins constrained 
to have the same CNO abundances relative to H.
The abundances are introduced into the fit via a linear scale factor of 
the metal ionic column densities predicted by the solar metallicity
grid of models.  
This is possible since all the absorbers considered here are optically thin to
the EUV ionizing radiation (i.e., the \heii\ ionization edge), thus
the total ionic column densities scale simply with the abundances of their
parent element.
In each velocity bin, there are measurements of $n_{ion}$ ionic
column densities (up to four, depending on whether there is contamination
or saturation present); the summation of these over velocity 
gives the total number of model constraints in each component.
This gives 2 $\times n_v +$ 3 free parameters, to be modeled with
$\Sigma_{i=1,n_v} n_{ion;i}$ constraints.
The best-fit solution is then the set of velocity-dependent $U$, $N_H$ values
and the C, N, and O abundances that minimize $\chi^2$ from:
\begin{equation}
\chi^2 = \Sigma_i \Sigma_j [\frac{N_{obs;j,i} - N_{mod;j}(U,N_H) f_j}{\sigma_{j,i}}]^2,
\end{equation}
where the measured column density of the $j^{th}$ ion in the $i^{th}$ velocity bin 
$N_{obs;j,i}$ is compared with the model column densities from the grid $N_{mod;j}$,
$f_j$ is the scale factor relative to solar abundances, and $\sigma_{j,i}$
the measurement uncertainty.  For bins that are consistent with being saturated 
within noise limits (adopted to be $\tau \geq$3), we set the upper $\sigma_{j,i}$ 
to a large value to effectively treat these as lower limits to the column densities.

\begin{figure}
\includegraphics[width=8cm]{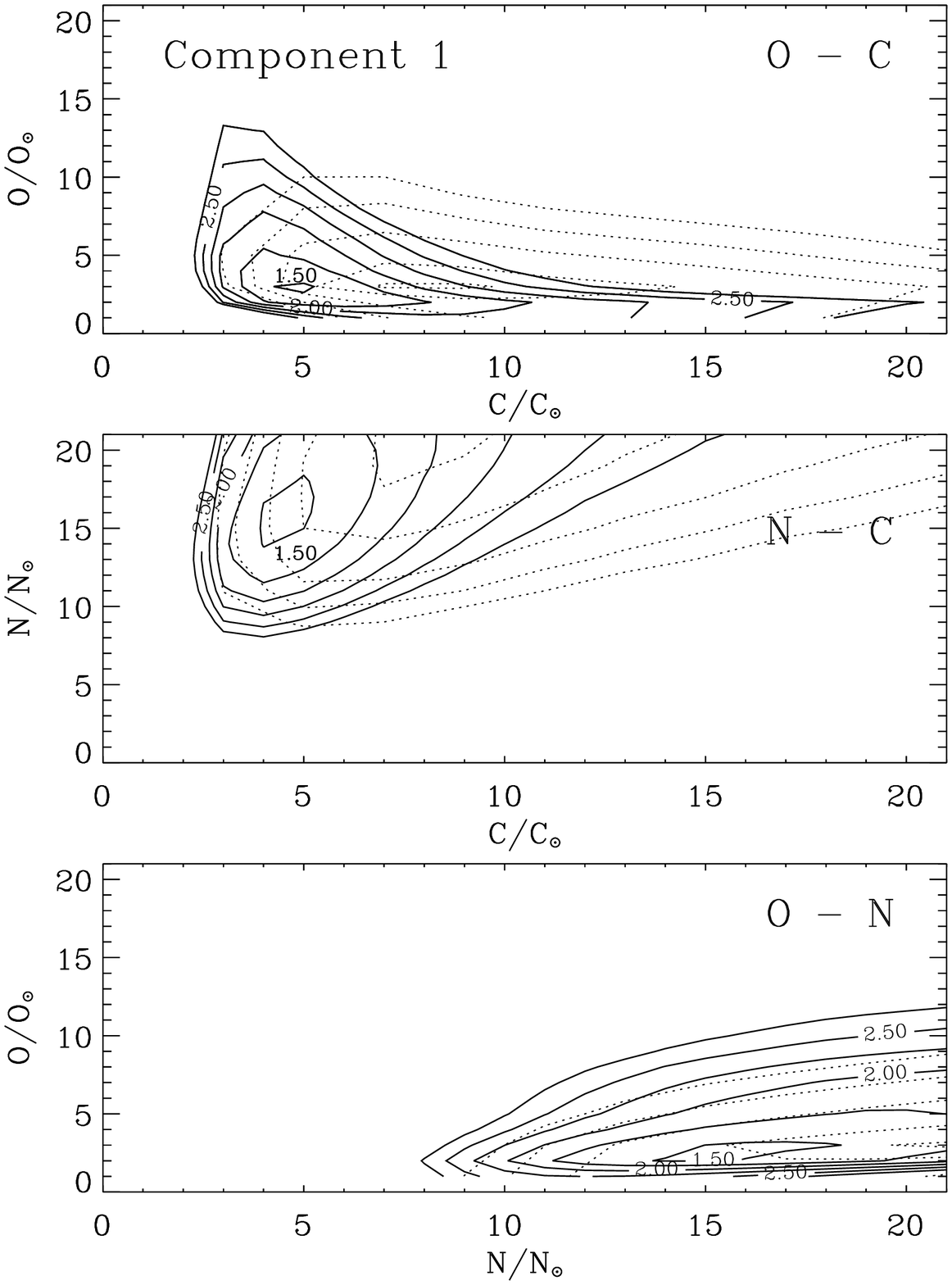}
\vspace*{0.35 in}
\caption{C, N, O abundance solutions from velocity-dependent photoionization
modeling for component 1.  The 2-D contour projections of the 3-D reduced $\chi^2$ 
volume are shown for each C-N-O plane. For each 2-D projection, the abundance
of the third, perpendicular element was held fixed at its best-fit value.
Solid contours show results for solutions to the column densities derived using
the global-fitting to all lines; dotted contours show results for column densities
derived individually for each ion pair, as described in the text.\label{fig5}}
\end{figure}

\begin{figure}
\includegraphics[width=8cm]{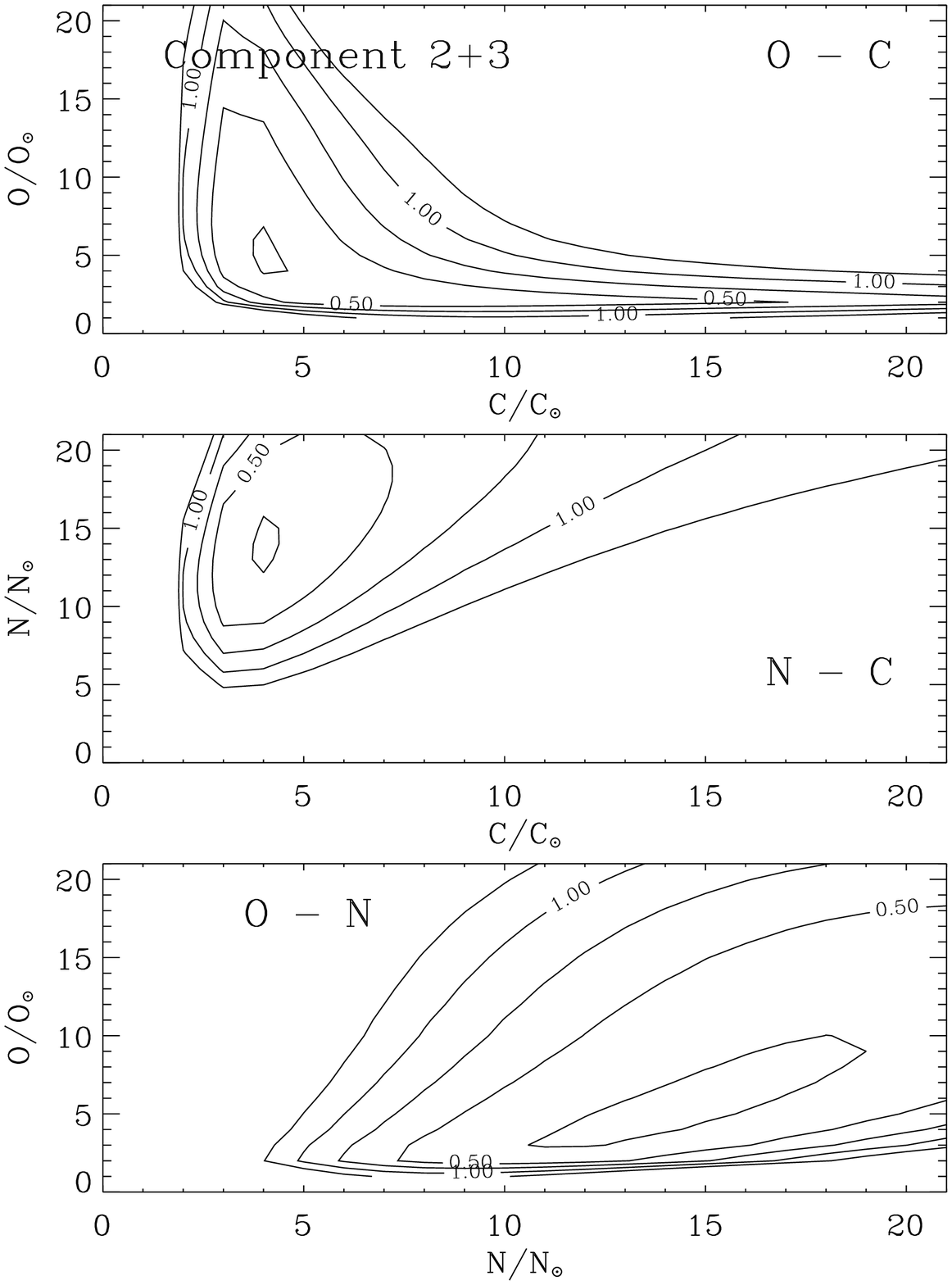}
\vspace*{0.35 in}
\caption{Same as Figure 5, showing the results for components 2+3 from modeling the
global-fit column densities. \label{fig6}}
\end{figure}

\subsection{Abundance Solutions}

    We have fit each of components 1--3 individually with this method,
as well as the combined absorption from all components (i.e., constraining
all kinematic components to have the same abundances).
The analysis indicates super-solar abundances in all kinematic components,
with generally consistent results for each element in each component.
The solutions for each component individually are 
[C/H]=0.6-0.7, [N/H]=1.1-1.2, [O/H]=0.5-0.7,
given as the log of the abundance ratio to hydrogen relative to solar values.
Thus C and O are enhanced by a factor of $\approx$ 3 -- 5 relative to solar
values, with additional overabundance of N consistent with $Z^2$ scaling 
indicative of enhanced secondary production in massive stars 
 \citep[e.g.,][and references therein]{hama99}.
These results can be compared with the study by 
\citet{peti99}, who found qualitative evidence for enhanced 
[N/C], but no strong indication of supersolar C and O abundances based on their 
modeling.
Figures 5 and 6 show the abundance solutions graphically for the independent 
fits to component 1 and components 2+3, respectively.
This shows the 2-D (contour) projections of the 3-D $\chi^2$ volume for the 
CNO abundances. The minimum reduced $\chi^2$ value for the component 1 fit
was 1.5, and for components 2+3 $\approx$0.25.
The low reduced $\chi^2$ value for components 2 and 3 may indicate the uncertainties
derived in our column density fitting are overly conservative for these 
components.  
For each 2-D $\chi^2$ contour projection, the abundance of the other element  
was held fixed at its best-fit value.

\begin{figure}[location=t]
\includegraphics[angle=90,width=8.5cm]{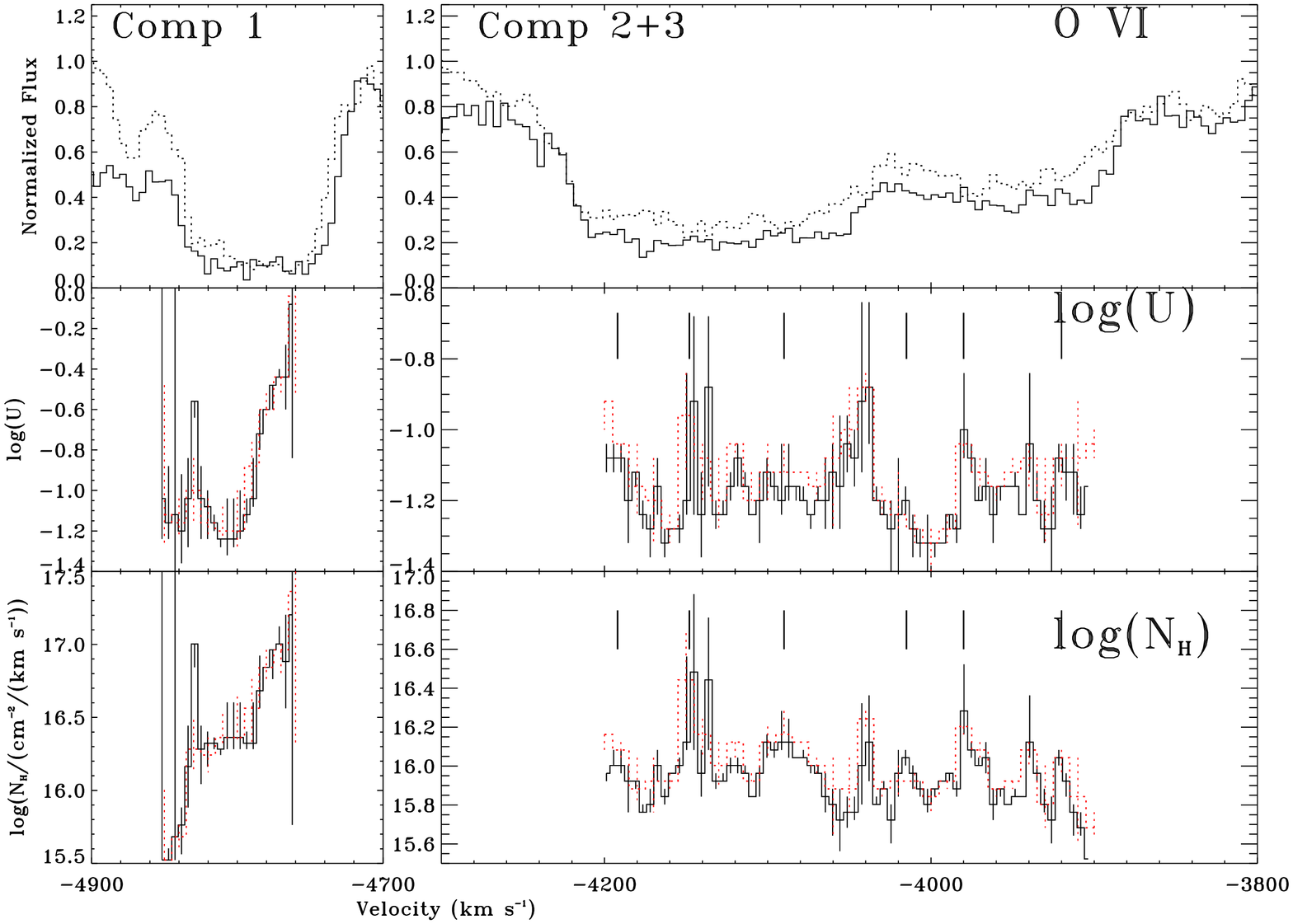}
\vspace*{0.1 in}
\caption{Velocity-dependent photoionization model solutions. The best-fit
solutions are shown for the ionization parameter (middle panel) and total H column
density (bottom), with metal abundances shown in figures 5 and 6. Black, solid 
histograms show the solutions derived at the full resolution of UVES; red, dotted
histrograms show solutions where the data were binned by $\times$2. 
Velocities of the narrow subpeaks identified in the optical depth 
profiles (Figure 3) are indicated with tickmarks.
The \ovi\ doublet profiles are shown in the top panel for reference.\label{fig7}}
\end{figure}

\subsection{Velocity-Dependent $U - N_H$ Solution}

    Figure 7 shows the solutions to $U$ and $N_H$ as a function of outflow velocity.  
The error bars represent values giving twice the minimum $\chi^2$ value from 
equation 3 with the abundances fixed at their best-fit values.  
Component 1 is seen to have fairly constant ionization structure ($log(U) \approx-$1.2)
over the blue-wing, and through the core of the component. 
The total hydrogen column density increases from the outer wing, and levels off to a
constant value over the central $\Delta v \approx$50~km~s$^{-1}$ in the core of the
profile.  In the red-wing, the total column and ionization increase sharply towards
lower outflow velocity.
This can be seen in the absorption profiles, where \ovi\ 
remains strong (saturated), while the other lines decrease.
At $v =-$4770~km~s$^{-1}$, the ionization parameter has increased to 
log($U$)$=-$0.4 and $N_H\approx$10$^{17}$~cm$^{-2}$, and
by $v =-$4750~km~s$^{-1}$, there are only lower limits on these parameters due
to the very weak absorption in the lower ionization lines.
This may be due to the superposition of two distinct physical components, or
could represent real physical velocity-dependent structure in the flow.
For components 2 and 3, the physical conditions in the absorbers are not strongly
velocity-dependent, with $log(U) \approx-$1.3 -- $-$1.0.
The total H column density does exhibit evidence for similar narrow substructure 
as the individual ionic column densities (\S 3), with $\approx$30\% fluctuations.
To test if this structure is an artificat of the solutions, in which strong 
fluctuations occur due to effects of large optical depths (near the saturation limit), we have 
also computed
models after binning the initial spectra by a factor of two. These results
are plotted as grey, dotted histograms on Figure 7 for comparison, showing that 
much of the narrow structure is present in these smoothed solutions.

\begin{figure}[location=t]
\includegraphics[width=8.5cm]{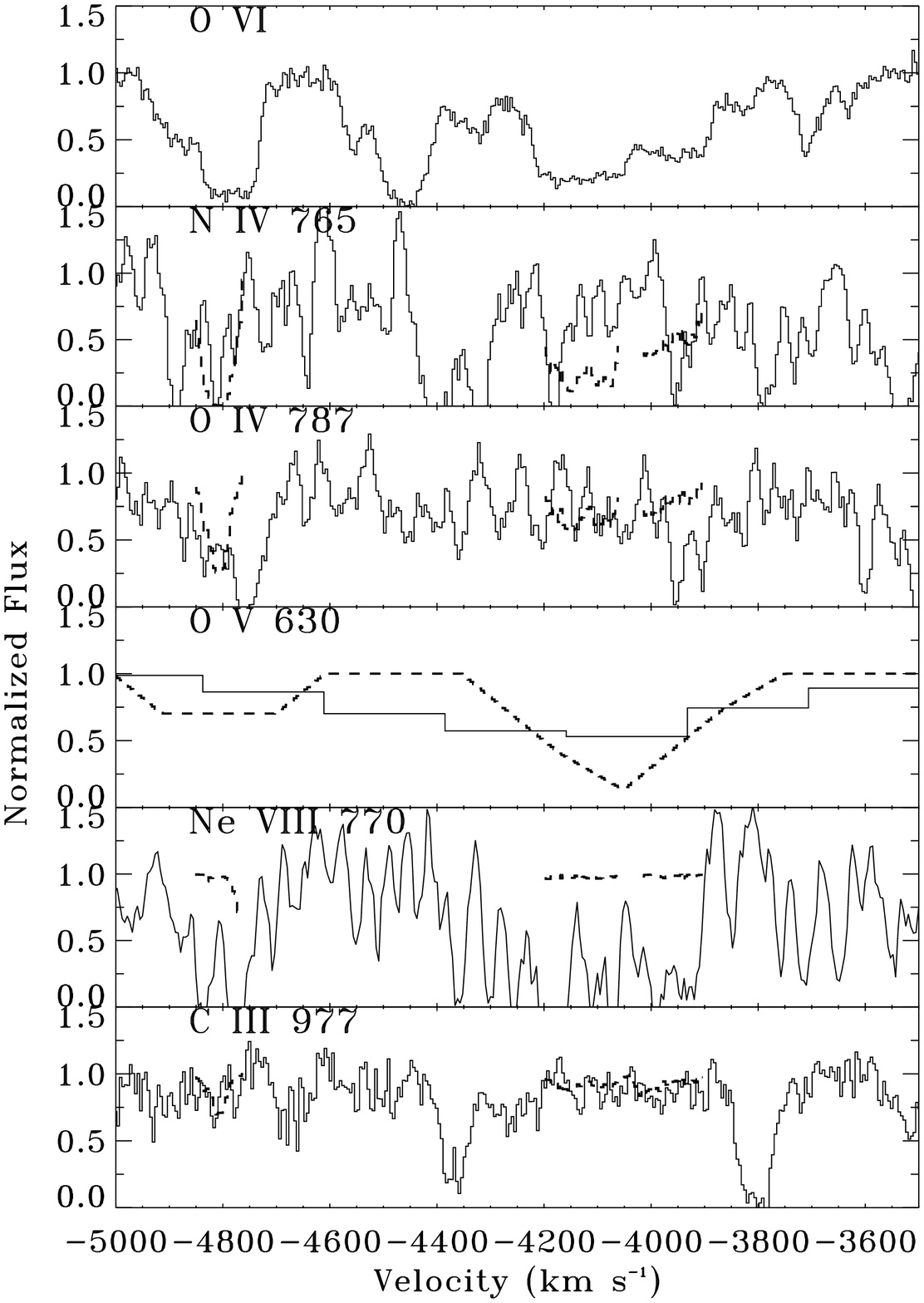}
\vspace*{0.35 in}
\caption{Model absorption profiles for other observed lines in J2233.
Synthetic profiles (heavy dashed lines) for lines not used in the photoionization
modeling are compared with their observed profiles.
They were derived using the predicted ionic column densities from
the photoionization model solutions and the covering factor solution.
Lines from the {\it HST}/STIS E230M spectrum (\niv,
\oiv, and \neviii) were smoothed over 5 bins; the model
\ov\  profile was binned to approximate the G140L resolution.
The \ciii\  spectrum is from the VLT/UVES observation.
The VLT/UVES spectrum of \ovi\  is shown at the top for reference.\label{fig8}}
\end{figure}

  We have generated synthetic absorption profiles for other lines observed in J2233 
using the predicted ionic column densities from the photoionization model solutions 
and the covering factor solutions.
These are shown in Figure 8 for key lines in the {\it HST}/STIS spectra 
and \ciii\ $\lambda$977 in the VLT/UVES spectrum.
Given the limited S/N in the STIS E230M spectrum 
(\niv\ $\lambda$765, \oiv\ $\lambda$787, 
and \neviii\ $\lambda$770) 
and the low-resolution of the \ov\ $\lambda$630 observation (G140L),
it is difficult to derive strong conclusions from the {\it HST} observations;
however, it is clear that the \neviii\ is far underestimated by
these models, requiring a higher ionization component as demonstrated in the
analysis by \citet{peti99}.
A similar situation was found in the case of QSO PG 0946+301 from a detailed
analysis of the highly constrained BAL by \citet{arav01a} - an additional
high-ionization absorber coincident in velocity with the other UV lines 
was needed to reproduce the observed \neviii\  absorption.
The other lines in J2233 are consistent with the observations, with the possible
exception of \niv\ $\lambda$765 in component 2 which may be overestimated.
The model predictions for \ciii\ $\lambda$977 are also seen to be consistent with
the observations. Additionally, absorption from the Si IV $\lambda$1393,1403 
doublet is predicted to be negligible in these models and consistent with
the tight upper limits on the Si IV column densities from the UVES spectrum (see 
Figure 1). We note that for the parameters derived here, these UV absorption
systems will produce negligible X-ray warm absorber features.

\subsection{High-Ionization Components 4 -- 6}

   Here we derive constraints on components 4 -- 6.
With only \ovi\ detectable in the UVES spectrum, there are not enough 
constraints to solve for the individual covering factors. However, comparison 
of the depths of the doublet members reveal the weak component 4 just
redward of component 3 has low covering factor. Though uncertainties 
from the doublet solution are rather large for
weak features like this \citep[see][]{gabe05a}, it requires 
that the continuum source is only partially covered (i.e., even if the 
BLR is fully unocculted), with $C_c$ at least as low as the value
derived for component 3.
The effective covering factors for components 5 and 6 are larger,
consistent with $C \approx$ 0.8 -- 1.
We note that these solutions should be taken with caution due to possible
contamination with the Ly$\alpha$ forest, and with no other lines
for confirmation.

   To constrain the physical conditions in these absorbers, we measured
\ovi\ column densities using the doublet solution \citep{hama97},
and derived upper limits on \hi, \nv, and \civ.
This gives lower limits on $U$, $N_H$ via photoionization modeling.
The ionization in all three components is substantially higher than
components 1 -- 3.
Adopting the abundances derived above for components 1 -- 3 implies
log($U$) $\geq$ $-$0.25, 0, and $-$0.4 
and log($N_H$) $\geq$ 17.7, 18.4, 17 for components 4 -- 6 respectively,
but the values could be substantially greater.
Upper limits on $U$, $N_H$ come from the \hi\   upper limit,
giving log($U$) $\leq$ 0.5,  log($N_H$) $\leq$ 19.7 for each component.
Assuming the features in the \neviii\  STIS spectrum are real for
these components, the implied column densities are consistent with the
upper range.

\subsection{Implications of More Complex Absorption Scenarios}

   As described in \S 3.1, our solution to the absorption
parameters ($\tau$,$C$) involved some simplifying 
assumptions. Specifically, we assumed: (1) the covering factors 
are identical for all ions (separately for the continuum and BLR 
sources), and (2) the absorption column density for each ionic
species is uniformly distributed over the occulted
fraction of the background emission (i.e., it is a perfect 
step-function in our sightline 
to the emission regions).  We explore the implications
of these assumptions in more detail below.

\begin{center}
{\it Ion-Dependent Covering Factors}
\end{center}

  The assumption that all ions have the same covering factors
was invoked in our global fitting to increase the number of fitting 
constraints so that the individual continuum and BLR 
covering factors could be solved.  
The results from this fitting matched the observed profiles 
well overall (see Figure 4), indicating the assumption 
is not too far off.
However, there are some discrepant regions in the
fit which may signify a more complex absorption-emission geometry 
in the J2233 outflow;  
primarily, the absorption strength in the \civ\ blue (red) 
doublet member is systematically underestimated (overestimated) over 
the core region of component 1 and in some regions 
of component 2. 

   To explore this further, we have computed column densities and
covering factors for each doublet/Ly$\alpha$-Ly$\beta$ pair 
individually using insight from our global fitting.  
Specifically, we fixed the continuum
covering factor at the average value for each kinematic
component from the global
fitting ($C_c =$ 1, 0.92, 0.75 for components 1, 2, 
and 3 respectively; see Figure 2), and then solved
for $\tau$ and $C_l$ separately for each ion line
pair by applying $\chi^2$ analysis to equation 1. 
We believe this is the best approach to explore ion-dependent
covering factors since the BLR is 
much larger than the UV continuum source, 
thus differences in covering factors 
between ions would most likely be due to affects of
BLR coverage.
For 1$\sigma$ uncertainties in these parameters, we adopted 
the maximum offsets from the best-fit solutions which give
$\chi^2 =$2, since there are two lines being fit. 
Solutions can only be obtained at velocities where both members
of the line pair are uncontaminated with other absorption. 
Bins that are saturated (we adopted $\tau \geq$ 3) have only
lower limits on their column densities.
In Figure 3, we plot the column density profiles derived
from this analysis.
The primary differences of these solutions are that the
\civ\ column density is somewhat larger in component 1 (and
equivalently $C_l$ is lower), by up to $\approx$50\%, 
and the associated 1$\sigma$ uncertainties
for all ions are generally larger than in the global-fit solutions.  
Figure 3 shows that most of the narrow substructure in the optical depth 
profiles in components 2 and 3 described in \S 3.2 is still 
present in the individual ion solutions.

   To test the potential affect of ion-dependent covering factors 
on our photoionization modeling results,
we did the analysis described in \S 4.1 using the column densities
derived individually for each ion in component 1 
(most of components 2 and 3 have no solutions for either \nv\ or 
H I due to contamination; see Figure 3).
The results for the abundances are presented in Figure 5, showing the
$\chi^2$ values as dotted contours. 
The carbon and nitrogen abundances derived from this model are 
about a factor of 1.5 times greater than the solutions using the
global-fit column densities, whereas the oxygen abundance is 
similar. We note that since \ovi\ is saturated over a large
portion of the component 1 profile and thus has only lower limits on
the column density, this may lead to an underestimate
of the oxygen abundance.

\begin{center}
{\it Inhomogeneous Absorption}
\end{center}

 Another assumption we have made in our $\tau - C$
measurements is that the absorber is a perfectly  
homogeneous slab, with uniform column densities spanning 
the occulted portion of the background emission.  
More complicated distributions of outflow material, with
a continous range of optical depths distributed with different
covering factors (i.e., an inhomogeneous absorber), were 
first explored by de Kool et al. (2001).  
More recently,   \citet[][hereafter AK05]{arav05} and 
\citet[][hereafter SH06]{sabr06}  
presented detailed analyses of inhomogeneous absorber (IA) distributions 
and their implications for observed absorption troughs.
Results from these studies indicate inhomogeneous absorption 
will not have a large affect on our modeling results and 
conclusions for the outflow in J2233, as described below.

    In AK05, it was shown that for lines from the same
ionic level that have different optical depths (e.g., 
doublet line pairs or Lyman series lines), 
cases where the observed residual fluxes in the two line troughs
are finite and equal in high S/N spectra (i.e., non-black 
saturation in the traditional partial coverage models), 
cannot be explained straightforwardly by IA models. 
Instead, they require a sharp edge in the distribution of 
absorber material, consistent with the homogeneous 
partial coverage model.
Specifically, using simple functional forms for the absorber material 
(power-law and Gaussian distributions), 
AK05 showed that ``shallow" distributions 
(i.e., power-laws with small spectral indices or Gaussians 
with relatively large $b$ parameters) cannot produce
finite but equal residual fluxes for any conditions, and steep distributions 
require unrealistically high values for peak optical
depths (see their Figures 2 and 3).  
The more general calculations of SH06, which allowed 
non-zero minimum bounds on the optical depth distribution,
alleviated the peak $\tau$ requirements somewhat for cases
of low residual fluxes.
However, even for these calculations, 
there is no way to obtain an absorption profile 
for a single line pair that gives equal residual fluxes
at some velocities but different fluxes at others 
unless the functional form of the distribution of
material is velocity-dependent across the component 
(e.g., see the models
with $C_f =$0.9 for the HPC models in Table 1 of SH06).
This is precisely the case observed in the troughs of J2233,
for example in \ion{O}{6} and \ion{N}{5} in component 1.
Such a situation seems physically implausible.
Thus we conclude that the IA models as presented in 
the detailed studies of AK05 and SH06 do not provide
a natural interpretation for the J2233 outflow.

  Finally, we note SH06 have shown that even for cases where
inhomogeneous absorption does apply, the mean optical depths
derived by assuming homogeneous partial coverage are very similar
to the mean values for IA distributions for most cases (they diverge
somewhat for strongly peaked distributions). 
Thus, the solutions from the homogeneous partial covering analysis
are generally representative of the inhomogenous distributions 
(for cases that are not saturated), once the mean optical
depths are accounted for by averaging over the unocculted 
and occulted regions.
It has yet to be shown, however, the implications for photoionization
models which are done assuming a slab of uniform column density.

\section{Implications for the Global Nature of the J2233 AGN Outflow}

   Here we synthesize the above results on the kinematics,
geometry, ionization, and abundances to
explore the global nature of the outflow in J2233.

\subsection{Chemical Enrichment}

    Determining abundances in quasar outflow systems serves as
a probe of the chemical evolution of the universe,
with implications for the timing and mechanisms for nucleosynthesis
and distribution of metals on galactic and intergalactic scales.
Analysis of intrinsic narrow absorption-line systems provides the
best means for probing the abundances of QSO environments:
BAL lines are typically heavily saturated and blended preventing
determination of ionic column densities, whereas observed emission-lines
sample the integrated flux over a large spatial region, and
thus likely sample a large range of physical conditions.
Evidence for near-solar or super-solar abundances has 
been found in associated absorbers in several other high redshift 
($z \sim$ 2--4) QSOs \citep[see references in][]{hama99,dodo04}, 
although this is often not definitive due to limited quality
of spectra. Additionally, linking these results to QSO outflows is uncertain in many
cases since they lack other properties that would indicate the
absorbers are intrinsic to the AGN.

   As shown in \S 4.2 and Figures 5 and 6, all
three outflow components in J2233 with available fitting constraints
have supersolar abundances, with enhanced N enrichment:  
$Z \approx$ 3--5 $Z_{\sun}$ for C and O, and N/H scaling roughly as $Z^2$.
The similarity in results derived independently in each kinematic component
gives strong support for this finding.
Additionally, the enhanced N abundance relative to C and O 
supports the general finding of high abundances in the absorbers,
since secondary CNO nucleosynthesis becomes  
important after the system has reached near solar metallicity, 
based on observations and chemical enrichment simulations 
\citep[e.g.,][]{hama93,vila93}.
The advanced chemical processing implied by our detailed fitting 
indicates the host galaxy of J2233 underwent vigorous
star formation at an early epoch, before $z=$2.2.
Given the finding of high metallicity in associated absorbers in other
bright, distant QSOs, this supports the evolutionary scenario 
in which massive galactic nuclei undergo rapid, extensive star formation at early times, 
with most chemical enrichment occurring before at least $z \gtrsim$ 2--3 
\citep[][and references therein]{hama99}. 
A subsequent or coeval QSO phase may then distribute the enriched gas to
the host galaxy, and possibly IGM, via an outflow.
The finding of distances of AGN outflows from the nucleus on kpc scales or greater  
\citep{deko01,hama01} supports this general model that QSO winds
can seed their environments with metals.

\subsection{Kinematic-Geometric-Ionization Structure}

   The detailed results of our absorption fitting and modeling
reveal correlations in the kinematic, geometric, and ionization 
structure in the J2233 outflow.
For example, our analysis in \S 3 (Figure 3) shows the covering factors 
of the individual kinematic components with low-ionization absorption 
are correlated with outflow velocity.  
To summarize, the highest velocity absorber, component 1, 
fully occults the continuum source and covers $\approx$ 60\% 
of the BLR, component 2 is consistent with full continuum coverage 
but little or no BLR coverage, while component 3 only partially 
occults the continuum source ($\approx$70\%).  

     One interpretation of this trend is that the absorption region size increases
with outflow velocity, at least as projected on the
sky against the background AGN emission (i.e., in the transverse dimension, 
$X_T$).  
{\it If} the absorbers in our sightline to the nuclear 
emission have a uniform 
density filling their entire volumes (i.e., filling factor $f =$1), 
e.g., if each component is a single ``cloud" or continuous flow region, 
then the expression relating the H column density to the
volume density and radial thickness of the absorber is 
$N_H = \vy{n}{H} \Delta R$. 
Since the integrated $N_H$ for the low-ionization regions 
are similar for components 2 and 3 and somewhat greater for 
component 1, if the radial dimensions of the absorbers follow 
the same trend as the transverse sizes in the scenario described
above (e.g., for a cloud-like geometry), then the volume densities 
decrease with outflow velocity.
Furthermore, the similarity in ionization parameter for the 
low-ionization material in each component (\S 4, Figure 7) implies their 
$n_H R^{2}$ products are similar via the expression for $U$ given in \S 4.1.  
Combining these results would imply a correlation between 
outflow velocity and $R$ for the components in J2233 if all
of the above assumptions apply.  
Qualitatively, this result is consistent with the predictions 
from dynamical models which have continuous acceleration of
a radial outflow.  Another interpretation of the $C - v_{out}$ 
correlation is that outflow regions with higher velocity have moved
further across our sightline to the emission sources.

\subsubsection{Geometric Constraints}

  The distance of the absorber from the central
ionizing source ($R$) has important implications for assessing the
total energy and mass flux associated with the outflow.
The most direct and accurate way of determining $R$ is by 
measurement of the volume density in the absorbers, $\vy{n}{H}$, 
from excited level absorption \citep[e.g.,][]{deko01,deko02b,deko02c,gabe05b}, 
but the presence of these diagnostic lines in AGN outflows is rare.
They are not detectable in the spectrum of 
J2233 within the noise limits; this is due to the relatively low-ionization
of the species giving excited level UV absorption (e.g., \siii* $\lambda$1260,
\cii* $\lambda$1336, \ciii* $\lambda$1175) compared with the ionization state 
of the outflow.  However limits on $R$ and the geometrical structure of
the outflow can still be explored based on indirect arguments involving 
other measurable quantities.

   For example, if the filling factors of the absorbers are unity 
as described above, then combining the expressions for $N_H$ and
the ionization parameter (\S 4.1) following \citet{turn93} gives 
the geometrical parameters $R$ and $\Delta R$ in terms of
the parameters derived from photoionization models, the H-ionizing photon
flux, and constants:  
\begin{equation}
\frac{\Delta R}{R^2} = \frac{N_H U 4 \pi c}{Q}.
\end{equation}
The photon flux can be estimated from the luminosity distance to J2233 
\citep[20 Gpc for $H_o =$ 71 km~s$^{-1}$~Mpc$^{-1}$,
$\Omega_m=$0.27, $\Omega_{\Lambda}=$0.73;][]{sper03} 
and integration of our model SED over the ionizing continuum, giving
$Q =$3$\times$10$^{57}$~s$^{-1}$.

  We can use the covering factors derived for the BLR (\S 3)
and the size of the BLR to constrain the projected transverse size ($X_T$)
of the absorption regions in our sightline to the background AGN 
emission.  Using the relation 
\citep[$R_{BLR} \propto L_{Bol}^{0.5}$; e.g.,][]{wand99}, 
this gives $X_T \geq$ 1~pc for component~1 ($C_l \approx$0.5),
and smaller for components 2 and 3 ($C_l \approx$0).
If the radial sizes of the absorption regions are similar to these limits on
the transverse sizes, $\Delta R \approx X_T$, consistent with a cloud-like geometry, 
then equation 4 implies $R \approx$ 50~kpc if the filling factor of
the absorbing medium is unity.
This large distance seems improbable, requiring the presence of extremely small structure
relative to its distance from the nucleus, with just the right line-of-sight geometry
to partially cover the AGN emission. However, similar results have been implied in 
analysis of other QSO outflows \citep[e.g.,][]{hama01}.
Alternatively, if the outflow has a thin shell geometry ($\Delta R \ll X_T$),
it could be considerably closer in to the nucleus; this model was adopted by 
\citet{stee05}, for example, to explain the X-ray absorption in
NGC 5548.  Another alternative is that the volume filling factor of the absorption 
material in our line-of-sight is small (i.e., it is highly clumpy). In this
case, the effective radial thickness of the absorber is proportional to 
$f^{1/3} \Delta R$, placing the outflow closer to the nucleus for a given 
$\Delta R$ : $X_T$ ratio.
We note that very low volume filling factors appear to be a generic feature 
of QSO outflows seen in the UV, and its implications and requirements on dynamical
models was recognized early on in the study of these systems 
\citep[e.g.,][]{weym85,arav94}.

\subsubsection{Narrow Kinematic Substructure and Velocity-Dependent Ionization}

  An interesting feature of the outflow in J2233 is the prevalence
of narrow kinematic structure seen in the absorbers.
This is directly apparent in the clustering of distinct
kinematic components 2 -- 6, which combine at adjacent 
velocities to give continuous absorption over $\Delta v >$ 500~km~s$^{-1}$.
Also, the detailed fitting and modeling in \S 3 and 4 
revealed the relatively broad component 2 and 3 absorbers
are comprised of a series of narrow peaks in column density 
($\Delta v \approx$ 25 -- 50~km~s$^{-1}$).
Further, component 1 shows strong velocity-dependent ionization 
differences: the red-wing is of significantly higher ionization,
with saturated \ovi\  absorption and weak or negligible absorption
from lower-ionization species.
This may be interpreted as the superposition of two physically
distinct components, partially overlapping in velocity space.

   Similar clustering in velocity of multiple, relatively 
narrow components is commonly seen in QSO outflows.  
For example, many BALQSOs show narrow substructure in some of the 
weaker lines coinciding in velocity with or adjacent to the broad 
troughs \citep[e.g.,][]{voit93,arav99,deko01,arav01b}.
In a recent multi-epoch study of QSO HS 1603+3820, \citet{misa05}  
found a cluster of partially overlapping narrow components had become
deeper and more blended, appearing almost as a BAL trough.
These results indicate we are seeing a number of distinct absorption regions 
in our line-of-sight to the AGN emission, 
either different stream lines associated with continuous 
outflows or distinct clouds.
Perhaps there is a continuous distribution of absorption widths associated
with AGN outflows, ranging from the narrow components seen in J2233 to the classical
BAL troughs, with the observed width related to the details of the outflow mechanism or
orientation of our sightline to the nucleus.
The clustering in velocity suggests the distinct regions are
related, possibly determined by the details of their origin or driving mechanism.
It is interesting that the low-ionization component 1 and 3 absorbers in J2233 
both have higher ionization absorption at adjacent,
lower outflow velocities.
Similar structure was seen in the QSO 2359-1241 absorber, 
and was taken as evidence for ionization stratification in a disk
wind by \citet{arav01b}.
In contrast, \citet{voit93} found some objects show the opposite trend,
with low-ionization absorption coinciding in velocity with the low-velocity
region of BAL troughs.

\section{Summary}

  We have presented a detailed analysis of the intrinsic
UV absorption in the central HDFS target, QSO J2233-606.
Our analysis is based on a high-resolution 
($R \approx$ 30,000 - 50,000), high S/N ($\sim$ 25 -- 50) 
spectrum obtained with VLT/UVES.
This spectrum samples the cluster of intrinsic absorption 
systems outflowing from the AGN at radial 
velocities $v \approx -$5000 -- 3800 km~s$^{-1}$ in 
the lithium-like CNO doublets and \hi\  Lyman series lines,
which serve as key diagnostics to determine the covering factors,
column densities, ionization state, and abundances of the outflow.

   We fit the absorption troughs as a function of radial velocity 
using a geometric model which assumes independent line-of-sight
covering factors for the individual background emission sources
(continuum and emission-line) and the same covering factors for all ions. 
The solution reveals:

\begin{enumerate} 
\item There is a trend of increasing covering factor for
components with greater outflow velocity:
component 1 at $v \approx-$4950~km~s$^{-1}$ occults the entire continuum source
and $\approx$55\% of the BLR; component 2 at $v \approx-$4150~km~s$^{-1}$
covers the continuum source, but little or none of the BLR; 
component 3 at $v \approx-$3950~km~s$^{-1}$ only partially covers
the continuum source ($\approx$70\%).

\item Narrow kinematic substructure in the ionic column densities 
is prevalent.  The red-wing of component 1 is highly ionized, having  
saturated \ovi\ and weak absorption from lower ionization species compared
with the core of this component.
Components 2 and 3 break down into a series of subpeaks in $\tau$, seen
consistently in the solutions to different ions.
\end{enumerate}

   We performed detailed photoionization modeling to determine 
the metal abundances and physical conditions in the absorbers.
Ionic column densities were simultaneously fit for all velocities 
of each component by constraining 
the abundances in each velocity bin to be the same.
This technique greatly increased the number of modeling constraints,
allowing a fit to the individual elemental abundances of C, N, and O
relative to H, as well as the velocity-dependent ionization parameter
and total column density.  Our analysis shows:

\begin{enumerate} 
\item All absorption components have supersolar metal abundances,
with [C/H] and [O/H] $\approx$0.5 -- 0.9, and [N/H] $\approx$ 1.2.  
This is consistent with the Z$^2$ scaling of N predicted from secondary
CNO nucleosynthesis.
Consistent results are found for independent fits to each component 
individually.
 
\item The lowest ionization material in each of components 1 -- 3 has
similar ionization (log($U$) $\approx-$1.2). Higher ionization gas is
required at the same velocities to account for the \neviii\  absorption,
as shown also by \citet{peti99}.
The redwing of component 1, and components 4 -- 6 which lie just redward of
of component 3, have ionization parameters
at least 5 -- 10 times greater than the low-ionization component 1 -- 3 gas.
\end{enumerate}

\bigskip

We are grateful to the ESO Archive for providing public 
access to data obtained with VLT/UVES. 
Support for this work was provided by NASA through grants 
AR-09536.08-A and NAG5-12867.
We thank G. Ferland 
for making his photoionization code Cloudy publicly 
available and C. Markwardt for providing access to his
optimization software. We also thank the referee for
insightful comments that lead to an improved paper.


\clearpage

\end{document}